\documentclass[10pt,twoside]{article}

\usepackage{times}
\usepackage[utf8]{inputenc}
\usepackage[T1]{fontenc}
\usepackage{graphicx}

\usepackage{multicol}
\usepackage{epstopdf}
\usepackage{booktabs}
\usepackage{amssymb}
\usepackage{amsmath}
\usepackage{multirow,hhline}
\usepackage{dirtytalk}

\usepackage{jep2022}
\usepackage[french]{babel}

\title{Anonymisation de parole par quantification vectorielle}

\author{Pierre Champion\up{1,2}\quad Denis Jouvet\up{1}\quad Anthony Larcher\up{2}\\
  {\small
    (1) Université de Lorraine, CNRS, Inria, LORIA, F-54000 Nancy, France \\ 
    (2) LIUM, Le Mans Université, Avenue Olivier Messiaen, 72085 LE MANS CEDEX 9, France \\
    \texttt{
      \{pierre.champion, denis.jouvet\}@inria.fr, anthony.larcher@univ-lemans.fr \\ 
}}}

\begin{document}
\maketitle

\resume{
L'utilisation de la reconnaissance de parole se répand de plus en plus dans les assistants virtuels.
Cependant, les signaux de parole contiennent de nombreuses informations sensibles telles que l'identité du locuteur, ce qui soulève des préoccupations quant à la protection des données personnelles.
Les expériences présentées montrent que les représentations extraites par les couches profondes des réseaux de reconnaissance de la parole contiennent cette information.
Dans cet article, nous cherchons à produire une représentation anonyme tout en préservant les performances de reconnaissance de la parole.
Dans ce but, nous proposons d'utiliser la quantification vectorielle pour contraindre l'espace de représentation, et inciter le réseau à supprimer l'identité du locuteur.
Le choix de la taille du dictionnaire de quantification permet d'ajuster le compromis entre l'utilité (reconnaissance de la parole) et le respect de la vie privée (masquage de l'identité du locuteur).
}

\abstract{Privacy-Preserving Speech Representation Learning using Vector Quantization}{
With the popularity of virtual assistants (e.g., Siri, Alexa), the use of speech recognition is now becoming more and more widespread.
However, speech signals contain a lot of sensitive information, such as the speaker's identity, which raises privacy concerns.
The presented experiments show that the representations extracted by the deep layers of speech recognition networks contain speaker information.
This paper aims to produce an anonymous representation while preserving speech recognition performance.
To this end, we propose to use vector quantization to constrain the representation space and induce the network to suppress the speaker identity.
The choice of the quantization dictionary size allows  to configure the trade-off between utility (speech recognition) and privacy (speaker identity concealment). 
}

\motsClefs
  {Anonymisation de la parole, Assistants vocaux, Reconnaissance du locuteur, Reconnaissance de parole}
  {Speech Anonymization, Voice Assistants, Speaker Recognition, Speech Recognition}

\vspace{-1em}
\section{Introduction}
\vspace{-1em}
Avec l'essor des assistants vocaux, de plus en plus d'objets connectés sont déployés chez les consommateurs.
Ces assistants ont besoin d'une connexion internet et de serveurs centralisés pour fonctionner. 
Les signaux de parole de l'utilisateur sont envoyés à ces serveurs pour bénéficier d'une expérience confortable et accessible en permanence.
Sur les serveurs, les fournisseurs de service font appel à des systèmes de reconnaissance automatique de parole et de compréhension du langage naturel afin de répondre à la demande de l'utilisateur.
Cependant, les signaux de parole contiennent de nombreuses informations relatives au locuteur, on y retrouve des attributs sensibles comme le genre du locuteur, son l'identité, son âge, ses sentiments, ses émotions, etc.
Ces attributs sensibles peuvent être extraits et utilisés à des fins malveillantes. 
Cette collecte excessive, et sans précédent, de signaux de parole sert à établir des profils d'utilisateurs complets et à construire de très grands jeux de données, nécessaires pour enrichir et améliorer les modèles de reconnaissance et de compréhension.
Ce transfert global des données vers les fournisseurs de services soulève de sérieuses questions à propos de la protection de la vie privée.
Récemment, des systèmes de reconnaissance de parole embarqués ont été proposés affin de résoudre cette problématique.
Cependant, les performances de ces systèmes sont encore restreintes dans les environnements peu favorables (c’est-à-dire, environnements bruyants, parole réverbérée, forts accents, etc.).
La collecte de grands corpus de parole représentatifs des utilisateurs réels et des diverses conditions d'utilisation est nécessaire pour améliorer les performances. 
Mais cela doit être effectué tout en préservant la vie privée des utilisateurs, ce qui signifie au moins garder l'identité du locuteur privée.

Dans l'approche proposée, un encodeur réside sur chaque objet connecté et effectue des calculs locaux pour créer une représentation anonymisée de la parole.
Ce processus de calcul défini par \cite{hybrid_privacy_framework} est adapté pour les assistants vocaux.
Jusqu'à présent, les travaux suivants s'y sont inscrits :
dans \cite{mohanPrivacyPreservingAdversarialRepresentation2019_reality_adversarial}, les auteurs emploient une méthode d'apprentissage antagoniste pour supprimer l'identité du locuteur dans un réseau de reconnaissance de la parole.
Cependant, leur approche a eu un fable impact, le système de vérification locuteur n'a pas vu ses performances significativement dégradées.
Dans \cite{private_wake_word}, les auteurs cherchent à créer une représentation capable de détecter des mots de réveil sans être capable de décoder le contenu linguistique.
Dans \cite{kmean_asr_privacy_configuratble}, les auteurs ont étudié la discrétisation de la parole dans de multiples systèmes de reconnaissance de la parole afin de minimiser l'inférence de plusieurs attributs sensibles (comme le locuteur,  l'émotion, le genre).
Finalement, dans le Challenge Voice Privacy (VPC) 2020 \cite{tomashenkoVoicePrivacy2020Challenge}, un protocole dédié et des métriques ont été proposés afin d'évaluer différentes méthodes d'anonymisation du locateur.

Dans cet article, notre travail est similaire à ceux de \cite{mohanPrivacyPreservingAdversarialRepresentation2019_reality_adversarial,kmean_asr_privacy_configuratble} ou nous nous concentrons sur la création d'une représentation anonymisée, où l’objectif est d'envoyer au fournisseur de services uniquement les informations qui lui sont nécessaires pour un bon fonctionnement du service.
Dans le cas considéré des assistants vocaux, l'information relative au contenu linguistique doit être gardée alors que celle relative aux locuteurs doit être supprimée. 
L'encodeur effectuant l'anonymisation étudiée dans cet article est basé sur un système de reconnaissance de la parole.
La représentation est extraite au niveau de la couche \textit{bottleneck} du réseau.
Ce type de représentation a pour but de compresser l'information afin qu’elle soit efficiente.
Dans le cas d'un système de reconnaissance de la parole, les \textit{bottlenecks} sont supposés encoder l'information du contenu linguistique, et ce, en étant invariants aux locuteurs. 
En utilisant le protocole d'évaluation du VPC, nous avons observé que les \textit{bottlenecks} n'encodent pas uniquement l'information linguistique, le locuteur peut être lui aussi identifié à un degré élevé.
Afin de mieux supprimer l'information relative au locuteur (donc améliorer l'anonymisation), nous avons introduit l'utilisation de la quantification vectorielle au niveau de la couche \textit{bottleneck} du réseau de reconnaissance de la parole.
La quantification vectorielle consiste en l'approximation d'un vecteur continu par un autre vecteur de même dimension, mais ce dernier appartenant à un ensemble fini de vecteurs \cite{vector_quantization}.
La quantification vectorielle est fréquemment utilisée dans la compression de données avec pertes.
Dans notre cadre d'utilisation, la quantification vectorielle permet d'imposer une contrainte sur le la couche \textit{bottleneck}.
Cette contrainte incite le réseau de reconnaissance de parole à encoder l'information du contenu linguistique dans un ensemble fini de vecteurs.
De ce fait, les autres informations relatives au locuteur se retrouvent moins encodées par manque de capacité d'encodage. 
Nos contributions sont les suivantes.
Premièrement, nous évaluons à quelle hauteur l'information du locuteur est présente dans un \textit{bottleneck} d'un système de reconnaissance de la parole.
Deuxièmement, nous étudions l'impact que la quantification vectorielle a sur les performances de reconnaissance de parole et du locuteur.
Troisièmement, nous montrons que les \textit{bottlenecks} peuvent être utilisés pour générer un signal de parole audible permettant une potentielle annotation et réapprentissage du modèle de reconnaissance de la parole.

La structure du reste du document est la suivante.
Dans la section \ref{sec:hybrid_framework}, nous décrivons le cadre de travail et le modèle proposé pour anonymiser la parole. La section \ref{sec:experiments} explique le dispositif expérimental et présente nos résultats. Enfin, nous concluons et discutons des travaux futurs dans la section \ref{sec:conclusion}.

\vspace{-1em}
\section{Processus de calcul hybride avec des calculs locaux et mutualisés}
\vspace{-1em}
\label{sec:hybrid_framework}


Dans cette section, nous présentons le cadre de travail hybride proposé par \cite{hybrid_privacy_framework} permettant d'effectuer des calculs locaux et d'autres mutualisés tout en respectant la vie privée des utilisateurs.
L'objectif de ce processus de calcul est de partager une représentation de parole avec un fournisseur de service, mais ce, en anonymisant les données de parole au niveau du périphérique avant de les partager.
Dans le contexte des assistants vocaux, la représentation anonymisée doit être riche en information relative au contenu linguistique tout en empêchant l'exposition d'informations sensibles qui pourrait potentiellement révéler des informations privées de l'utilisateur.
Dans nos expériences, nous nous focalisons sur l’identité du locuteur, et considérons que cette information doit être supprimée.
Dans ce processus de calcul hybride, la tâche compliquée est de concevoir l'encodeur qui extrait la représentation anonymisée, car le codage ou la modification du signal de parole peut nuire au bon fonctionnement de la tâche de reconnaissance de la parole.
Dans la section suivante, nous décrivons l'architecture de l'encodeur utilisé pour anonymiser la parole. 

\vspace{-1em}
\subsection{Présentation du modèle}
\vspace{-0.5em}

De par leurs fonctions d'apprentissage, les modèles acoustiques utilisés dans les systèmes de reconnaissance de parole cherchent à encoder l'information du contenu linguistique (par exemple via la classification temporelle des phonèmes).
Ces modèles sont souvent conçus pour être invariants au locuteur dans le but de proposer les mêmes performances de reconnaissance à tout utilisateur.
C'est pour ces raisons que nous avons choisi d'utiliser comme encodeur un modèle acoustique.

Nous utilisons une architecture \textit{time delay neural network factorized} (\textit{TDNN-F}) introduite par \cite{Povey2018SemiOrthogonal_TDNNF}.
Elle est utilisée dans le cadre d'un système de reconnaissance de la parole hybride \textit{Hidden Markov Model - Deep Neural Network (HMM-DNN)} \cite{KaldiPovey}.
Cette architecture a été reconnue comme l'une des plus efficientes dans un récent classement comparant les performances des modèles par rapport aux exigences matérielles \cite{Performance_vs_hardware_asr}.
L'architecture \textit{TDNN-F} est donc appropriée pour l'utilisation embarquée nécessaire au fonctionnement du processus hybride avec des calculs locaux et d'autres effectuées par un serveur centraliser.

La fonction d'objectif \textit{Lattice-Free Maximum Mutual Information (LF-MMI)} \cite{lfmmi} est utilisée afin de réaliser un entraînement discriminatif des séquences.
La fonction \textit{MMI} traditionnelle vise à maximiser la probabilité postérieure :
\begin{equation}
\begin{aligned}
\mathcal{L}_{mmi}(\lambda) =\sum_{r=1}^{R} \log P_{\lambda}\left(S_{r} \mid O_{r}\right)
&=\sum_{r=1}^{R} \log \frac{P_{\lambda}\left(O_{r} \mid S_{r}\right) P\left(S_{r}\right) }{\sum_{S}P_{\lambda}\left(O_{r} \mid S\right) P\left(S\right)}
\end{aligned}
\label{eq:lfmmi}
\end{equation}

où $\lambda$ est l'ensemble des paramètres du réseau de neurones, $R$ est le nombre total de segments d'apprentissage,
$S_{r}$ est la transcription correcte du $r^{eme}$ segment de parole $O_{r}$,
$P(S)$ est la probabilité du modèle de langage pour la phrase $S$.
La distribution $P\left(S\right)$ est considérée comme fixe, et est estimée avec un modèle de langage à partir des transcriptions d'entraînement.
Le numérateur indique la vraisemblance de la prédiction pour une séquence de mots de référence, tandis que le dénominateur indique la vraisemblance totale de la prédiction pour toutes les séquences de mots possibles, ce qui équivaut à la somme sur toutes les séquences de mots possibles estimées par le modèle acoustique et le modèle de langage.
Le numérateur encode les caractéristiques de supervision et il est spécifique à chaque segment, tandis que le dénominateur encode toutes les séquences de mots possibles et il est identique pour tous les segments.
Cette fonction de coût est optimisée en maximisant le numérateur et en minimisant le dénominateur.
\textit{MMI} maximise la log-vraisemblance conditionnelle des probabilités globalement normalisées des transcriptions correctes.

\begin{figure}[t] 
\begin{center} 
\includegraphics[width=1.0\linewidth]{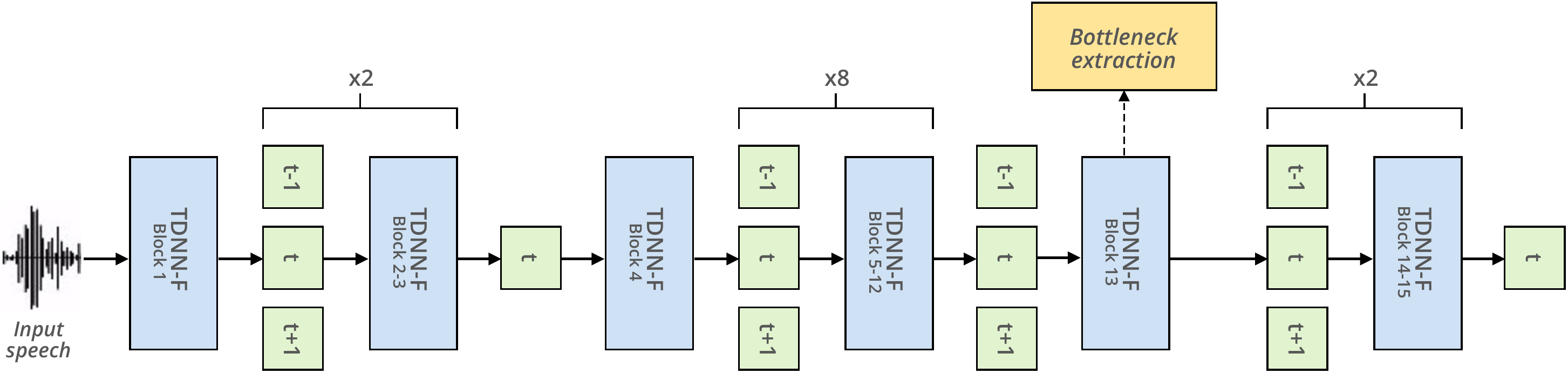}
\end{center} 
\caption{
Architecture du modèle \textit{TDNN-F}, totalisant 15 couches. 
Les \textit{bottleneck} sont extraits à partir de la 13e couche.
} \label{image:tdnnf} \
\vspace{-1em}
\vspace{-1em}
\end{figure}

Dans l'objectif d'obtenir une représentation anonymisée de la parole, nous extrayons des \textit{bottlenecks} de faibles dimensions ($D$ = 256 dimensions) depuis une couche profonde du réseau (la 13ème couche sur les 15 du réseau, cf.: figure \ref{image:tdnnf}).
Il a été observé par \cite{adiReverseGradientNot2019,mohanPrivacyPreservingAdversarialRepresentation2019_reality_adversarial,whatdoesanetworkhears} que ce type de représentation encode principalement l'information relative au contenu linguistique et supprime une partie de l'information de l'identité locuteur.

\vspace{-1em}
\subsection{Introduction à la quantification vectorielle pour l'anonymisation}
\vspace{-0.5em}
Afin d'améliorer l'anonymisation, nous proposons de contraindre la couche du réseau de neurones produisant les \textit{bottlenecks} en ajoutant une couche de quantification vectorielle.
La quantification vectorielle consiste en l'approximation d'un vecteur continu par un autre vecteur de même dimension, mais ce dernier appartenant à un ensemble fini de vecteurs \cite{vector_quantization}, ces vecteurs sont dénommés vecteurs prototypes.
Dans la tâche d'apprentissage non supervisé de représentation discriminante via l'utilisation d'auto-encodeur, il a été observé que les vecteurs prototypes appris suite à une quantification vectorielle représentent principalement l'information relative aux phonèmes \cite{neural_disctre_vq,Unsupervised_speech_rep_vq,one-shot-vc-vector-quant}.

L'application de la quantification vectorielle dans un modèle acoustique a pour but d'inciter le modèle à supprimer l'information du locuteur, car la quantification vectorielle réduit la capacité d'encodage du réseau.
Comparée aux tâches non supervisées, la fonction de coût d'un modèle acoustique impose explicitement que l'information phonétique soit encodée dans le \textit{bottleneck}, de ce fait, nous pouvons appliquer une contrainte élevée en réduisant le nombre de vecteurs prototypes.

Étant donné la séquence audio d'entrée $s=\left(s_{1}, s_{2}, \ldots, s_{T}\right)$ de longueur $T$, l'encodeur \textit{TDNN-F} produit les \textit{bottlenecks} $h(s) = \left(h_{1}, h_{2}, \ldots, h_{j}\right)$ de longueur $J$ ($J < T$ dû au sous-échantillonnage effectué par l'encodeur) où $h_{j} \in \mathbb{R}^{D}$ pour chaque pas temporel $t$, et $D$ est la dimensionnalité de la représentation latente.

La couche de quantification vectorielle prend en entrée la séquence de vecteurs continus $h(s)$ et remplace chaque $h_{j} \in h(s)$ par un prototype du dictionnaire apprenable $E=\left\{e_{1}, e_{2}, \ldots, e_{V}\right\}$ de taille $V$, chaque $e_{i} \in \mathbb{R}^{D}$.
\begin{equation}
q(s)=\underset{e_i}{\arg \min }\left\|h(s)-e_{i}\right\|_{2}^{2}
\end{equation}
Le vecteur $h_{j}$ est remplacé par son vecteur prototype $e_{v}$ le plus proche en termes de distance euclidienne.
Puisque la quantification est non différentiable (à cause de l'opération $\arg \min$), sa dérivée doit être approximée.
Pour ce faire, nous utilisons un \textit{straight-through estimator} \cite{strat_through_estimator} i.e.,$\frac{\partial \mathcal{L}}{\partial h(s)} \approx \frac{\partial \mathcal{L}}{\partial q(s)}$.
Les vecteurs prototype sont contraints de se rapprocher des vecteurs \textit{bottlenecks} qu'ils remplacent par l'ajout d'une fonction de coût auxiliaire:
\begin{equation}
\mathcal{L}_{vq} = \left\|\operatorname{sg}\left[h(s)\right]-q(s)\right\|_{2}^{2}
\end{equation}
où $\mathrm{sg}[\cdot]$ désigne l'opération bloquant la rétropropagation du gradient, donc la mise à jour des poids.
Cette opération est semblable à un k-means, mais appliquée à chaque minibatch pendant l'apprentissage, les prototypes du dictionnaire correspondant aux centroïdes d'un k-means.
Étant donné que les \textit{bottleneck} peuvent prendre n'importe quelle valeur, l'ajout d'une fonction de coût régularise l'encodeur a produire des \textit{bottlenecks} proches des prototypes afin que l'apprentissage de l'encodeur ne diverge pas de l'apprentissage du dictionnaire :
\begin{equation}
\mathcal{L}_{vq\_reg} = \|h(x)-\operatorname{sg}[q(x)]\|_{2}^{2}
\end{equation}

La fonction de coût du modèle acoustique peut être alors exprimée par la somme des fonctions \textit{mmi}, de quantification et de régularisation :
\begin{equation}
\mathcal{L}=\mathcal{L}_{mmi}+\mathcal{L}_{vq}+\lambda \mathcal{L}_{vq\_reg}
\end{equation}
où $\lambda$ désigne le coefficient du facteur de régularisation (nous avons utilisé $\lambda = 0.25$).
Pour mettre à jour les prototypes du dictionnaire, nous utilisons une moyenne mobile exponentielle (EMA) \cite{ema_vq}. 
EMA met à jour le dictionnaire $E$ indépendamment de l'optimiseur, l'apprentissage est donc plus robuste face aux différents choix d'optimiseurs et d'hyperparamètres (par example: le taux d'apprentissage, momentum).



\vspace{-1em}
\section{Expériences}
\vspace{-1em}
\label{sec:experiments}

\subsection{Jeux de données}
\vspace{-1em}
Nous avons utilisé le corpus LibriSpeech \cite{Librispeech} pour toutes nos expériences.
Les statistiques des jeux de données sont disponibles dans le tableau \ref{tab:data-train}.

\begin{table*}[htbp]

  \caption{Statistiques des jeux de données d'entraînement et de test.}\label{tab:data-train}
  \centering
  \begin{tabular}{l r r r r c }
      \toprule
   \multirow{2}{*}{{}} &   \multirow{2}{*}{{Taille}} & \multicolumn{3}{c}{{Nombre de locuteurs}} &  \multirow{2}{*}{
   \begin{minipage}[t]{0.14\textwidth}
   \centering
   {Nombre d'utterances}
   \end{minipage}
   } \\ 
   &  & {Femme} & {Homme} & {Total} & \\
       \midrule
  LibriSpeech: train-clean-100 & 	100h	& 125 &	126 &	251	& {~~28539}	\\ 
  LibriSpeech: train-clean-360 &    364h    & 439 & 482 &	921	& {104014} \\
  LibriSpeech: test-clean &     	5.4h    &  20 &  20 &	 40	& {~~~~2620} \\
  LibriSpeech: test-other  &     	5.1h    &  17 &  16 &	 33	& {~~~~2939} \\
  \bottomrule
  \end{tabular}
\end{table*}
Le sous-ensemble LibriSpeech train-clean-100 a été utilisé pour apprendre le modèle acoustique.
Les jeux de données utilisées pour évaluer les performances de reconnaissance de parole sont LibriSpeech test-clean et LibriSpeech test-other.

Le challenge Voice Privacy définit LibriSpeech train-clean-360 comme jeux d'apprentissage pour apprendre le système de vérification du locuteur.
Il est important de remarquer que ce jeu d'apprentissage ne propose pas une grande variabilité intralocuteur du aux longues sessions d'enregistrement des chapitres des livres audio.
Entraîner le système de vérification du locuteur sur les \textit{bottlenecks} d'un modèle acoustique n'est pas une tâche facile, car toute erreur de représentation effectuée par le modèle acoustique est propagée dans celui de vérification du locuteur.
Pour atténuer cet effet, nous avons appris le système de reconnaissance du locuteur sur la combinaison de train-clean-100 et train-clean-360.
Le modèle acoustique produit une très bonne représentation pour le sous-ensemble train-clean-100 (vu lors de l'entraînement du modèle acoustique), ce qui aide l'apprentissage du modèle de vérification du locuteur.

Conformément au challenge Voice Privacy, les performances en reconnaissance du locuteur ont été évaluées avec le jeu de donnée LibriSpeech test-clean.
Parmi les 40 locuteurs de LibriSpeech test-clean, 29 d'entre eux sont sélectionnés, pour chaque locuteur un sous-ensemble totalisant 1 min de parole (après détection d'activité vocale) a été sélectionné pour l'ensemble d'enrôlement et le reste a été utilisé pour l'ensemble de test.
Les nombres de test cible et imposteur sont détaillés dans le tableau \ref{tab:data-test}.

\vspace{-1em}
\begin{table*}[htbp]
  \caption{Nombre de test de vérification dans l'ensemble de donnée d'évaluation.}\label{tab:data-test}
  \vspace{0.5mm}
  \centering
  \begin{tabular}{l c r r c}
      \toprule
 {} & {Type} &  {Femme} & {Homme} & {Total}  \\
      \midrule
\multirow{2}{*}{Librispeech: test-clean} & Cible 	& 548	& 449	& {~~~997} \\ 
& Imposteur  	&	{11196} &	{9457} &	{20653} \\ 
  \bottomrule
  \end{tabular}
\end{table*}
\vspace{-1em}

\vspace{-1em}
\subsection{Métriques et évaluation}
\vspace{-0.5em}
\label{sec:metrics}
Pour évaluer les performances du système en matière d'anonymisation (\textit{capacité de dissimulation de l'identité du locuteur}) et d'utilité (\textit{capacité  à reconnaître le contenu linguistique}), deux systèmes et métriques sont utilisés.

Pour évaluer quantitativement la qualité de l'anonymisation, une architecture de vérification automatique du locuteur implémentée dans SideKit \cite{sidekit} est utilisée.
Il s'agit d'un système x-vecteur composé de cinq couches TDNN suivit d'une couche de \textit{statistics pooling}  \cite{snyder2018xvector}.
La fonction de coût utilisée pour l'apprentissage est la \textit{large margin softmax loss} \cite{aamlossArcMarginProduct}.
La métrique d'évaluation est le taux d'égale erreur (EER$_\%$), plus l'EER$_\%$ est élevé, mieux les locuteurs sont anonymisés.

Pour l'utilité, le système de reconnaissance de parole transcrit la parole depuis la représentation \textit{bottleneck}.
La mesure du taux d'erreurs mots (WER$_\%$) est utilisée pour évaluer dans quelle mesure les \textit{bottlenecks} encodent correctement l'information linguistique.
Plus le WER$_\%$ est faible, mieux le contenu linguistique est encodé.


\vspace{-1em}
\subsection{Résultats et discussions}
\vspace{-0.5em}
Le tableau \ref{tab:results} présente les résultats expérimentaux.
La première ligne présente les scores de vérification du locuteur et de reconnaissance de parole pour les jeux de données test-clean et test-other, sans introduction de quantification vectorielle.
Ces scores sont cohérents avec ceux reportés dans la littérature \cite{tomashenkoVoicePrivacy2020Challenge,pkwrap}.
Les résultats de vérification du locuteur montrent que la représentation \textit{bottleneck} d'un système de reconnaissance de parole de référence (No VQ) est capable de correctement discriminer les locuteurs. 
Il est à noter que pour cet exemple les femmes sont plus difficiles à différencier que les hommes,  c'est-à-dire : 9,3 EER$_\%$ pour les femmes et 4,2 EER$_\%$ pour les hommes. Le WER$_\%$ sur LibriSpeech test-clean est de 5,8, valeur utilisée comme référence pour les expériences suivantes.

En contraignant la représentation \textit{bottleneck} du réseau avec l'utilisation de quantification vectorielle, les performances de vérification du locuteur sont drastiquement réduites.
Le nombre $V$ de prototypes dans le dictionnaire de quantification contraint plus ou moins le modèle acoustique,
avec $V$ vecteurs prototypes l'information linguistique du signal est compressée dans un espace vectoriel discret de $V$ vecteurs prototypes. 
Plus le dictionnaire est petit, plus le réseau doit trouver une transformation efficace pour représenter l'information linguistique, ce qui laisse moins de place pour encoder l'information relative au locuteur. 
Ainsi avec $V$ = 16 les scores d'EER$_\%$ de vérification du locuteur sont de 30,0 pour les femmes et 32,4 pour les hommes valeurs plus élevées qu'avec $V$ = 1024 ou le réseau obtient 17,9 pour les femmes et 18,3 pour les hommes.
En comparaison avec le système de référence (No VQ), l'utilisation de la quantification vectorielle permet d'anonymiser les \textit{bottlenecks}. 
Plus la valeur de $V$ est petite, moins les \textit{bottlenecks} sont représentatifs du locuteur, permettant une meilleure anonymisation.

Cependant, les performances de reconnaissance de parole, mesurées en termes de WER$_\%$, sont elles aussi impactées par la taille du dictionnaire de quantification.
Avec $V$ = 16 le WER$_\%$ est de 15,9 sur LibriSpeech test-clean, dégradation très importante par rapport à la valeur de référence de 5,8. En augmentant le nombre de vecteurs prototype, le WER$_\%$ redescend. Pour $V$ = 1024 le WER$_\%$ est de 7,2.
Le tableau \ref{tab:results} présente aussi les scores de reconnaissance de parole sur le jeu de données LibriSpeech test-other.

\begin{table*}[t]
    \caption{Résultats de la reconnaissance vocale et de la vérification du locuteur en fonction du nombre de vecteurs prototypes dans le dictionnaire de quantification. La mesure de l'intervalle de confiance pour l'EER et le WER est effectuée avec un ré-échantillonnage \textit{bootstrap}.}
    \centering
    \begin{tabular}{ c c c c c c }
    \toprule
         \multicolumn{2}{c}{}   Nb vecteurs  & \multicolumn{2}{c}{EER$_\%$} & \multicolumn{2}{c}{WER$_\%$} \\
         \multicolumn{2}{c}{}  prototypes  & F & H & test-clean & test-other \\
    \midrule
                         & (No VQ)  &  9.3 \footnotesize{$\pm$0.5}   & 4.2 \footnotesize{$\pm$1.0}    &   5.8 \footnotesize{$\pm$0.3}  &   19.5 \footnotesize{$\pm$0.6}    \\
    \midrule
                                        & 16  & 30.0 \footnotesize{$\pm$2.1}   & 32.4 \footnotesize{$\pm$2.1 } & 15.9\footnotesize{ $\pm$0.5 }  & 42.5\footnotesize{ $\pm$0.8} \\

                                      & 32  & 25.6 \footnotesize{$\pm$2.1 }  & 27.3 \footnotesize{$\pm$1.9 } & 9.8 \footnotesize{$\pm$0.4  } & 31.4 \footnotesize{$\pm$0.8 }\\

             & 48  & 22.0 \footnotesize{$\pm$1.7 }  & 22.6 \footnotesize{$\pm$2.1 } & 8.7 \footnotesize{$\pm$0.4  } & 28.8 \footnotesize{$\pm$0.8 }\\

                                        & 128 & 22.0 \footnotesize{$\pm$1.8 }  & 22.8 \footnotesize{$\pm$2.0 } & 8.5 \footnotesize{$\pm$0.4  } & 28.5 \footnotesize{$\pm$0.8 }\\

                                        & 256 & 19.2 \footnotesize{$\pm$1.6 }  & 19.6 \footnotesize{$\pm$2.0 } & 7.6 \footnotesize{$\pm$0.3  } & 26.1 \footnotesize{$\pm$0.7 }\\

                                        & 512 & 19.6 \footnotesize{$\pm$1.6 }  & 19.2 \footnotesize{$\pm$2.0 } & 7.6 \footnotesize{$\pm$0.3  } & 25.4 \footnotesize{$\pm$0.7 }\\

                                        & 1024 & 17.9 \footnotesize{$\pm$1.6}   & 18.3\footnotesize{ $\pm$1.8}  & 7.2\footnotesize{ $\pm$0.3 }  & 24.7\footnotesize{ $\pm$0.7} \\

    \bottomrule
    \end{tabular}
    \label{tab:results}
    \vspace{-1em}
\end{table*}

Le compromis entre de bonnes performances en reconnaissance de parole et une bonne anonymisation est inhérent au problème de partage de données anonymisées (problème connu sous le nom de \say{\textit{privacy-utility tradeoff}} \cite{privacy-utility-tradeoff}). 
Dans notre cadre de travail, ce compromis est paramétrable et peut être ajusté au souhait de l'utilisateur ou du fournisseur de service via la taille $V$ du dictionnaire de quantification.
De manière générale, le tableau \ref{tab:results} montre que plus $V$ est faible, meilleure est l'anonymisation, mais cela est au prix d'une dégradation des performances en reconnaissance de parole.
Et, inversement, plus $V$ est grand, meilleure est la reconnaissance de parole au détriment d'une moins bonne anonymisation.


\vspace{-1em}

\section{Conclusion}
\vspace{-0.5em}
\label{sec:conclusion}
Dans cet article, nous avons appliqué le processus de calcul hybride respectueux de la vie privée,  qui décompose un modèle neuronal en deux parties, un encodeur qui génère une représentation anonyme sur l'appareil de l'utilisateur, et un décodeur qui utilise cette représentation anonyme pour effectuer des calculs mutualisés.
Nous avons étudié ce système dans le contexte des assistants vocaux.
Comme encodeur, un modèle acoustique \textit{TDNN-F} a été considéré, et nous avons montré ses limitations.
En utilisant le jeu de donnée du challenge Voice Privacy, nous avons mesuré que le locuteur peut être vérifié à la hauteur de 9,3$_\%$ d'EER pour les femmes et 4,2$_\%$ d'ERR pour les hommes dans un modèle \textit{TDNN-F} classique.
Nous avons proposé d'utiliser un algorithme de quantification vectorielle afin de contraindre l'espace de représentation, forçant ainsi le modèle acoustique à uniquement encoder l'information phonétique.
Cet algorithme est configurable en fonction de la taille du dictionnaire de quantification, ce qui permet d'ajuster le compromis entre de bonnes performances en reconnaissance de parole et une bonne anonymisation.
Par exemple, avec un dictionnaire de 128 vecteurs, le locuteur est dramatiquement moins vérifiable, 22,0$_\%$ d'EER pour les femmes et 22,8$_\%$ d'ERR pour les hommes ce qui correspond a un gain 232\%.
Mais ce gain en anonymisation impacte les performances de reconnaissance de parole, le WER augmente de 47\% (augmentation de 5,8$_\%$ à 8,5$_\%$ de WER).
Dans les prochains travaux, nous prévoyons de générer de la parole à partir de ces représentations anonymes et d'évaluer les performances en reconnaissance de parole et masquage d'identité du locuteur à partir de la parole générée.

\vspace{-1em}
\vspace{-0.5em}
\section*{Remerciements}
\vspace{-0.1em}
Ce travail a été réalisé avec le soutien de l'Agence nationale de la recherche française, dans le cadre du projet ANR DEEP-PRIVACY (18-CE23-0018) et de la Région Grand Est.

\vspace{-1em}
\bibliographystyle{jep2022}
\bibliography{biblio}

\end{document}